# DIGITAL SORTING PERTURBED LAGUERRE-GAUSSIAN BEAMS BY RADIAL NUMBERS VIA HIGH ORDER INTENSITY MOMENTS

A. Volyar, M. Bretsko, Ya. Akimova, Yu. Egorov

*V.I. Vernadsky Crimean Federal University, Vernadsky Prospekt, 4, Simferopol, 295007, Russia*

## 1. Introduction

The unique properties of the Laguerre-Gauss beams (LG) [1] attract the special attention of researchers from various fields of science and technology in connection with relative simplicity and elegance of their employment. The properties of LG beams, whose mathematical structure in polar coordinates is written in a non-normalized form as

$$\psi_{m,n}(r,\varphi,z) = \frac{1}{w}\left(\frac{r}{w}\right)^{|m|} L_n^{|m|}\left(2\frac{r^2}{w^2}\right)\exp(im\varphi + i\Gamma_{m,n})\exp\left(-\frac{r^2}{\sigma}\right), \quad (1)$$

where $r = \rho/w_0$ is a dimensionless radius, $\rho$ and $\varphi$ are the polar coordinates, are controlled by the beam parameters: $w_0$ stands for the beam waist at the initial plane $z=0$, $\sigma = 1 - iz/z_0$, $z_0 = kw_0^2/2$ denotes the Rayleigh length, $k$ is a wavenumber, $w = |\sigma|$,

$$\Gamma_{m,n} = (2n+m+1)\arctan(z/z_0) \quad (2)$$

stands for the Gouy phase while $m$ and $n$ are the azimuthal and radial indices, respectively.

The integer index $m = 0, \pm 1, \pm 2, \ldots$ in the exponential factor $\exp(im\varphi)$ acts the part of the vortex topological charge (TC) i.e. it sets the phase change by $m$ cycles of $2\pi$ in any closed circuit around the beam axis whereas the beam amplitude is zero. It is the topological charge $m$ that is responsible for the fact that the optical vortex carries the orbital angular momentum (OAM) first noticed by Allen and co-workers in Ref. [2]. Such an exclusive property of LG beams immediately found wide application from physical research (see e.g. optical Hall effect on a boundary surface [3,4] and the similar effect in uniaxial crystals [5,6]) to astronomy and medical optics: the microparticle manipulations [7,8] (see also [9] and references therein), in astronomy for light twisting around black holes [10] and telescopes for observation of new planets in distant stellar systems [11], in optical communication in free space [12] and fibers [13], in microscopy [14] and even in medical optics [15] for a fractal analyzing of the vortex beams scattered by a human skin.

Against the background of abundant publications on OAM studies (the azimuthal quantum number $m$), a completely different attention was paid to the role of a radial quantum number $n = 0,1,2,3,\ldots$ in LG beams. It was believed that the radial index of LG beams simply fixes a number of degenerate ring dislocations in vortex beams. However, in recent years the situation has changed dramatically. It turned out that radial numbers allow one to shape darkness knots and control their shape [16, 17]. The original physical interpretation of the radial quantum number was presented in recent articles [18-21]. In essence, based on the operator formalism, authors of Ref. [18,21,34] revealed that there is a hidden symmetry of the state rendering an additional degree of freedom which is not represented by the azimuthal mode number $m$ but connected with the radial number $n$ called the *"intrinsic hyperbolic momentum charge"*[19]. Using a radial quantum number increases the reliability of observing and employing the "mysterious" effect of entangled quantum states [20,22,24] and enables one to achieve the theoretical maximum data transmission capacity in free-space communications [23]. A new theoretical approach to the questioned problem of the radial quantum number of photons and LG vortex beams required improving adequate measurement techniques.

As back as in the late 80s of the last century, a computer hologram technique for sorting laser mode beams was proposed and developed by Soifer and Golub [25]. Their research was based on analogy with a glass prism or a diffraction grating that decomposes a polychromatic light into a spatial spectrum of monochromatic beams. Further development of this approach was reflected in articles [26-28] for LG vortex modus. The authors formed holographic gratings in such a way that the propagation direction of the diffracted beams was determined by the azimuthal and radial numbers of the vortex modes and formed a matrix in rectangular or axial symmetry at the photodetector. The original method of sorting vortex beams based on a mode conversion was recently described in Ref.[29]. The method requires three consecutive phase modulations followed by a single mode fiber [30[. The key point the method is conversion of $N$ input beams into $N$ orthogonal modes of an optical fiber i.e. transformation a mode basis into another mode basis that can be separately detected. Although this approach is characterized by low energy losses, its complexity requires careful adjustment in order to achieve sufficiently accurate measurement results. Also we cannot but noting articles on the vortex spectra measurements based on an optical geometric transformation [31], analyzing interference patterns [32], a collinear phase-shifting holography [33] and using controlled random materials [35], that also enables one to manipulate of a radial mode number.

As a special group, we took out techniques of sorting vortex modes by only a radial quantum number. The most popular approach for sorting modes by radial number relies on the dependence of the Gouy phase $\Gamma_{m,n}$ (2) on the propagation distance $z$ of LG beams and the Rayleigh length $z_0$ [36]. The measurement process uses either a Mach – Zehnder or Sagnac interferometer types. A lens system is placed in one of the interferometer arms allowing controlling the Gouy phase. The second arm is for the reference beam. The wave interference after a beam image processing enables one to sort LG beams by radial number. The measurement error is critical to any variations of the beam waist and wavefront curvature radii at the photodetector plane; therefore, the device requires careful adjustment. Modifications of this approach are used in various optical devices, both for communication purposes and for entangled quantum states (see e.g. Ref.[23,24,37,38]). So, for example, in the article [37], the authors use two sequentially located Mach – Zehnder interferometers. The first interferometer includes a set of refractive optical elements to manipulate the fractional Gouy phase by realizing the fractional Fourier transform [39] while the second one discriminates the modes based on the induced phase. The described approach allows one to sort the modes by both the orbital and radial quantum numbers. Is it possible to significantly simplify multi-staged optical systems, excluding interferometric devices, so as to analyze the mode composition of the combined vortex beam directly without complex optical transformations?

One of the possible approaches to solving this problem has been recently proposed and implemented in the articles [40-43]. The key idea of this technique is to use the intensity moments of higher-orders (see e.g. [44,45] and references therein) to analyze the vortex composition of the combined singular beams at their waist plane without interfering in the wavefront structure. The authors of Ref.[40] considered specifics of employing the intensity moments technique for measuring the vortex spectrum (squared amplitudes and mode phases) and using them to calculating the total OAM of the combined beam containing LG modes with different topological charges $m$ of the same signs ($m>0$ or $m<0$) but zero radial number $n=0$. This technique allowed analyzing structure of the vortex avalanche [41] caused by small local perturbations of the holographic grating. The use of the astigmatic transformation of the combined beam by means of a cylindrical lens [42] has expanded this technique to the beams with different signs of topological charges ($m>0$ and $m<0$), but a zero radial number. This made it possible to analyze the vortex spectrum and measure both OAM and informational entropy (Shannon entropy) [46] in a vortex beam perturbed by a sectorial aperture [43]. However, considered approach cannot be used for analyzing laser beams containing LG modes with different radial numbers $n \neq 0$, that can occur even in a simplest situation of aperturing vortex beams.

Let us assume that we were able to develop and implement a technique for sorting LG modes by radial number (to plot the mode spectrum). But if the initial beam contains many LG modes, then each component in the mode spectrum is degenerate. The more modes the initial beam contains, the higher the degeneracy number of each component in the mode spectrum of the perturbed combined beam. But a question arises: How to sort LG modes in the case of degeneracy? This problem has not been previously investigated. Thus, the purpose of our article is to develop and implement the intensity moments technique for the digital sorting of perturbed LG beams by radial numbers taking into account the mode degeneracy of the spectral components.

## 2. Preliminary remarks

The intensity distribution $\mathfrak{I}_{00}(r,\varphi) = \Psi(r,\varphi)\Psi^*(r,\varphi)$ of a monochromatic beam with a complex amplitude $\Psi(r,\varphi)$ subjected to an external perturbation contains complete information about both the initial state of the light field and the source of the perturbation. The key problem is how to extract this information from the beam at the experiment. In quantum optics, the elements of the optical density matrix (see e.g.[47,48]) are considered in terms of the photon number basis and the Wigner distribution function (see [49] and references therein). Applications of the Wigner function technique was analyzed in detail for simple optical systems in Ref.[50] and found widespread employment in image recognition systems [44]. As we mentioned above, the intensity moments are conveniently used for sorting vortex modes of a complex beam by topological charge (an azimuthal number). In this section, we touch upon the problem of sorting LG modes by radial number that arise as a result of a perturbation of both a single LG beam and an array of such beams.

We focus on the perturbation of a monochromatic LG beam by a shaped aperture that does not cause a change of the vortex modes phase at the perturbation plane described by a function of real variables $f(r,\varphi)$. We write the perturbed beam field in the orthogonal basis of LG modes (1) at the initial plane $z = 0$ in the form

$$\Psi_{m,n}(\rho,\varphi, z=0) = \sum_{s=-\infty}^{\infty}\sum_{k=0}^{\infty} C_{m,n,s,k} \rho^{|s|} L_k^{|s|}\left(2\frac{\rho^2}{w^2}\right) e^{is\varphi} \exp\left(-\frac{\rho^2}{w^2}\right), \qquad (3)$$

where the mode amplitudes are defined as follows

$$C_{m,n,s,k} = \int_{\mathbb{R}} \Psi_{m,n} \psi_{s,k}^* dS, \qquad (4)$$

and $\Psi_{m,n} = f(\rho,\varphi)\psi_{m,n}(\rho,\varphi)$.

Recently, we have examined the problem of digital sorting of vortex beams by topological charge $m$ in the case of a local perturbation of the holographic grating [40-42] and sector aperture [43], provided that the radial number $n$ does not change. In this article, we draw attention to sorting vortex modes by radial number $n$ provided that the perturbation does not affect the azimuthal number $(m = const)$ of the initial singular beam. Obviously, such a perturbation associated with truncating the beam by a circular hard-edged aperture. Note that this problem was repeatedly considered to solve applied problems, for example, to optimize a finite number of modes in the expansion (3) (see Ref.[51] and references therein), but the authors did not address the problem of mode sorting.

## 3. Digital sorting of the LG modes

Before discussing the complex problem of the mode sorting when a combined beam containing LG beams array is perturbed (and where the mode cross-talk between the vortex modes with the same radial numbers $k$ plays a leading part), we explore the simplest case of a single LG beam perturbation. The requirement for an aperture function $f(r,\varphi)$ not to excite vortex modes with different azimuthal numbers $m$ but gives rise to LG modes with variety of radial numbers is fulfilled, for example, in the case of a circular hard-edged aperture

$$f(r) = circ(r) = \begin{cases} 1, & 0 \leq r \leq R \\ 0, & r > R \end{cases}. \qquad (5)$$

Now the complex beam amplitude (3) can be reduced to

$$\Psi_{m.n}(r,\varphi) = \sum_{k=0}^{\infty} C_{m,n,k} r^{|m|} L_k^{|m|}(2r^2) e^{im\varphi} \exp(-r^2) \qquad (6)$$

while the mode amplitudes are

$$C_{m,n,k} = \int_0^R \Psi(r,\varphi)\psi_{m,k}^*(r,\psi) r\, dr \bigg/ \int_0^\infty |\psi_{m,k}(r,\psi)|^2 r\, dr \qquad (7)$$

or

$$C_{m,n,k} = \frac{k!}{(m+k)!} \sum_{j=0}^{k} (-1)^j \binom{m+k}{k-j} \frac{(m+1)_n}{2^j j! n!} \frac{1}{m+j+1} \times \\ \times 2R^{2(m+j+1)} {}_2F_2\left(m+n+1, m+j+1; m+1, m+j+2; -(2R)^2\right) \qquad (8)$$

where ${}_2F_2(...)$ stands for a hypergeometric function, $(m+1)_n$ is a Pochhammer symbol and we used the integral [52]

$$\int_0^x x^\lambda e^{-x} L_n^m(x)\, dx = \frac{(m+1)_n}{n!(\lambda+1)} x^{\lambda+1} \times \\ \times {}_2F_2(m+n+1, \lambda+1; m+1, \lambda+2; -x) \qquad (9)$$

together with the orthogonality condition and the associated Laguerre polynomial representation

$$L_k^m(x) = \sum_{j=0}^{k} (-1)^j \binom{m+k}{k-j} \frac{x^j}{j!} \qquad (10)$$

The mode amplitudes $C_{m,n,k}$ in Eq.(6) are specified by a triple of numbers $(m,n,k)$. The azimuthal number $m$ or the topological charge of the vortex beam in the factor $\exp(im\varphi)$ does not share in the summation of the modes in Eq. (6) and, therefore, the orbital angular momentum of the beam does not change under perturbation. The perturbation process is controlled by the rest of two numbers $(n,k)$ in such a way that to each initial radial number $n$ before the perturbation there corresponds a wide interval of radial numbers $k$ after action of the perturbation. It is convenient to represent the result of the perturbation in the form of the vortex spectrum $C_{m,n,k}^2(n,k)$ shown in Fig. 1. Small variations of the circular aperture radius $R$ lead to a sharp change in the vortex spectrum shape that can be seen by comparing the spectra in Fig.1a and b. In fact, the vortex spectra in Fig. 1 illustrate the redistribution of light intensity of the initial beam with a radial number $n$ between LG modes with different radial indices k, i.e. sorting modes by radial numbers $k$.

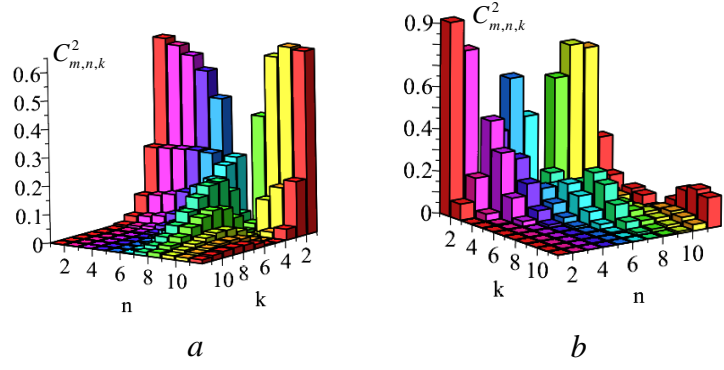

*Fig.1 The vortex spectra $C^2_{m,n,k}(n,k)$ for different aperture radii: (a) $R=1$, (b) $R=1.5$ and $m=3$. The squared amplitudes $C^2_{m,n,k}$ specify sorting the vortex modes by the radial number $k$ provided that the initial LG beam with the topological charge $m$ and radial number $n$ was perturbed by a circular aperture of the radius R. The sum of all $C^2_{m,n,k}(k)$ for the given number $n$ is equal to unity*

Now, the expansion of the complex amplitude of the perturbed paraxial beam in the $z=0$ plane can easily be extended to an arbitrary length $z$ by inserting the factor $1/w(z)$ in the radial coordinate $r$ and the Gouy phase $\Gamma_{m,k}$ (see Eq.(1))

$$\Psi_{m,n}(r,\varphi,z) = \frac{1}{\sigma(z)} \sum_{k=0}^{\infty} C_{m,n,k} \left(\frac{r}{w(z)}\right)^{|m|} L_k^{|m|}\left(2\frac{r^2}{w^2(z)}\right) e^{im\varphi + i\Gamma_{m,k}(z)} \exp\left(-\frac{r^2}{\sigma(z)}\right). \quad (11)$$

Thus, our main task is first experimentally to sort the LG modes by the radial number using the approach of high-order intensity moments, and then extend this technique for sorting the modes that arise when the array of LG beams are perturbed. The intensity moments are specified by the expression [42,44]

$$J_{p,q} = \int_{\mathbb{R}} M_{p,q}(r,\varphi) \Im\, r\, dS, \quad (12)$$

where $M_{p,q}(r,\varphi)$ stands for the moments function, $\Im_{m,n}(r,\varphi,z=0) = |\Psi_{m,n}(r,z=0)|^2$ is the beam intensity distribution, $p,q = 0,1,2,...$. Since the intensity distribution $\Im_{p,q}(r)$ is an axially symmetric function the moments function $M_{p,q}(r,\varphi)$ can be reduced to the form $M_{p,q}(r)$. The intensity distribution $\Im_{p,q}(r)$ can be written as

$$\Im_{m,n} = \sum_{k=0}^{\infty} C^2_{m,n,k} |\psi_{m,k}(r)|^2 + 2 \sum_{k,s=0}^{\infty} C_{m,n,k} C_{m,n,s} \psi_{m,n,k} \psi^*_{m,n,s}. \quad (13)$$

To choose the optimal form of the moments function $M_{p,q}$, in Eq.(12) it is necessary to find out the features of mode amplitudes $C_{m,n,k}$ in Eq. (13). As an example, Fig. 2 shows the distribution of mode amplitudes $C_{m,n,k}$ over the radial number $k$ of the LG beam perturbed by a circular aperture plotted according to Eq. (8). The vortex spectra in Fig. 1 and 2 enable us to make three important inferences. First, the LG mode intensities $C^2_{m,n,k}$ are quickly reduced with increasing radial number $k$. This means that the summation of the LG modes in Eq. (13) can be limited by the number $k_{\max} = N$ requiring, e.g., that the squared amplitudes be greater than $C^2_{m,n,N} \geq 0.01$. Secondly, the sign-changing spectrum of mode amplitudes $C_{m,n,k}(k)$ shown in Fig.2a means that for obtaining complete information about the perturbed beam, it is not sufficiently to measure only the squared amplitudes $C^2_{m,n,k}$ but also measure all cross terms in Eq. (13) to define the signs of the mode amplitudes. Third, the dependence curves 1, 2 and 3 in Fig. 2b show slow (curve 3) and fast oscillations (curves 2 and 1) of the squared amplitudes $C^2_{m,n,k}(R)$ on changing the radius $R$ of the circular hard-edged aperture for the initial LG beam with the topological charge $m=3$ and radial number $n=6$. Such fast oscillations of the mode intensities $C^2_{m,n,k}$ with small variations of the aperture radius $R$ impose strict restrictions on the choice of the moment function $M_{p,q}$. As we will show later, an unsuccessful choice of the moment function can lead to large errors in the digital mode sorting.

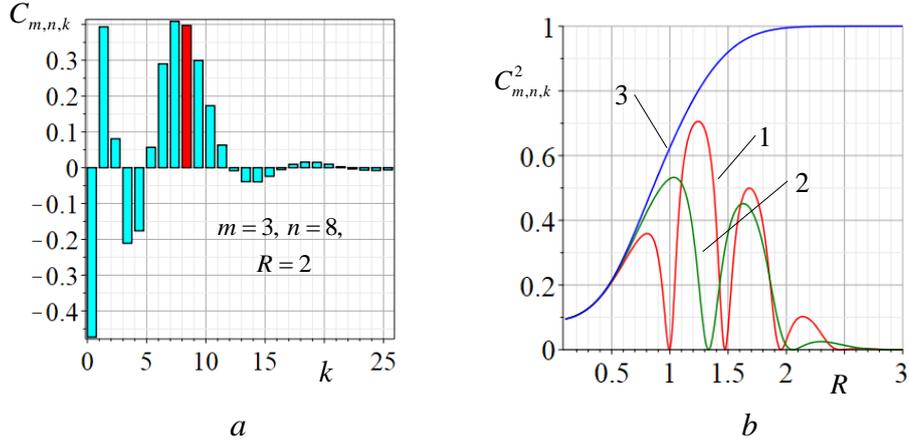

*Fig.2 (a) Distribution of the mode amplitudes $C_{m,n,k}$ by the radial numbers $k$ of the LG beam with numbers $m=3$, $n=8$ perturbed by the circular aperture of the radius $R=2$; (b) the $C^2_{m,n,k}(R)$ dependence of the hard-edged aperture radius $R$ for the mode radial numbers (1) $k=6$, (2) $k=3$, (3) $k=0$ provided that the initial LG beam with $m=3$, $n=6$ is perturbed.*

Note that the definition of the intensity moments in Eq. (12) can be treated as a system of linear equations with respect to the squared amplitudes $C^2_{m,n,k}$ and the cross terms $2C_{m,n,k}C_{m,n,s}$. Indeed, the terms on the right-hand side of equation (12) depend on the squared amplitudes and the cross terms and factors

$$j_{pq} = \int_0^\infty r^{2m+1} M_{p,q} L_k^m(2r^2) L_s^m(2r^2) e^{-2r^2} dr, \qquad (14)$$

that can be calculated. On the other hand, the product of the experimentally measured intensity distribution $\Im_{\exp}(r)$ and the moments function $M_{p,q}$, and therefore the entire integral (12) can be found experimentally in combination with the computer calculations provided that $\Im(r) \to \Im_{\exp}(r)$. The number of variables in the equations is specified by the number of squared amplitudes $X_p$ and cross terms $Y_p$. The number of the squared amplitudes is $N$ while the number of cross-amplitudes in (13) is defined as the 2-combination of a $N$-set equal to the binomial coefficient $N!/(2!(N-2)!)$ so that a total number of the linear equations is

$$N_{total} = N + \frac{N!}{2!(N-2)!}. \qquad (15)$$

Let us impose designations $X_P = C^2_{m,n,k}$, $Y_P = 2C_{m,n,k}C_{m,n,s}$ and form a column-vector $\mathbf{X} = (X_1\ X_2...X_N\ Y_{N+1}...Y_{N_{tot}})^T$ where the symbol $T$ indicates the transpose operation whereas the measured intensity moments are written as $\mathbf{J} = (J_1, J_2 ... J_{N_{tot}})^T$. Further, using Eq. (14), we form a matrix

$$\hat{K} = \begin{pmatrix} j_{11} & \cdots & j_{1s} & \cdots & j_{1N_{tot}} \\ \cdots & \cdots & \cdots & \cdots & \cdots \\ j_{s1} & \cdots & j_{ss} & \cdots & j_{sN_{tot}} \\ \cdots & \cdots & \cdots & \cdots & \cdots \\ j_{N_{tot}1} & \cdots & j_{N_{tot}s} & \cdots & j_{N_{tot}N_{tot}} \end{pmatrix} \qquad (16)$$

and at last we rewrite Eq.(12) in a form of the linear equations system

$$\mathbf{J} = \hat{K} \mathbf{X}. \qquad (17)$$

The terms obtained at the experiment are located on the right side of the equations, whereas the matrix elements are theoretically calculated. Then the squared amplitudes and cross amplitudes are found as

$$\mathbf{X} = \hat{K}^{-1} \mathbf{J}. \qquad (18)$$

However, the inverse matrix $\hat{K}^{-1}$ can be obtained only if the determinant of the matrix $\hat{K}$ is not equal to zero i.e. $\det \hat{K} \neq 0$. Thus, the necessary condition for the choice of the moments function is requirement $\det \hat{K} \neq 0$ while the sufficient condition is to minimize the measurement error of the components of the vector $\mathbf{X}$. The only way to evaluate such a measurement error is a computer simulation of the measurement process, comparing the predetermined mode amplitudes with those obtained at a computer experiment.

Let us consider the sorting modes process by the radial number using the above algorithm on the example of perturbation of an LG beam by a circular hard-edged aperture. First of all, it is necessary to choose the optimal form of the moment function $M_{p,q}$. Based on the form of the intensity distribution, it is logical to choose the function $M_{p,q}$ in the form of Laguerre functions, which corresponds to the orthogonal basis Laguerre polynomials for the intensity moments representation $M_{p,q} = L_q^p(2r^2)$. It was this approach that we used for sorting the vortex modes by the topological charge $m$ in the case of a non-degenerate vortex array ($m > 0$ or $m < 0$) [40]. However, the use of this kind of moment function turned out to be inapplicable in this case since the verification showed that $\det \hat{K} = 0$. We have chosen an alternative orthogonal basis for the representation in the form of trigonometric functions $M_{p,q} = \begin{pmatrix} \sin pr \\ \cos qr \end{pmatrix}$. Although now $\det \hat{K} \neq 0$, our computer simulation showed that in the general case $p \neq q$, the error of detecting the squared amplitudes can exceed 70% for different aperture radii $R$ and numbers $(m, n)$. The only combination of numbers in the form $p = q = 1, 3, 5, ...$ corresponded to a computer simulation error less than 1%.

An important element in sorting is measuring the cross amplitudes $Y_s$ and calculating the sign of each amplitude. The fact is that processing the combined vortex beam structure containing a set of LG modes requires analyzing not only the squared amplitude spectrum, but also signs of each mode. To specify the amplitude signs, it is sufficiently to know the amplitude sign of the initial unperturbed beam (e.g. $Y_n > 0$ term), and then find the sign of each mode using the chain of cross-amplitudes: $Y_0 \leftarrow ... \leftarrow Y_{n-1} \leftarrow Y_n \rightarrow Y_{n+1} \rightarrow ... \rightarrow Y_{N_{tot}}$.

In order to examine in detail the experimental process of the digital sorting LG modes by radial numbers, we choose three unperturbed LG beams with the same topological charge $m = 3$ but different radial numbers: $n = 0$, $n = 3$ and $n = 6$. First, we will implement sorting the LG modes for each perturbed beam, and then sorting the LG modes for superposition of all beams.

The digital sorting was carried out on the experimental setup whose sketch is shown in Fig. 3. A fundamental Gaussian ($TEM_{00}$ mode) beam emitted by a He-Ne laser ($\lambda = 0.6328$ mcm) passed through a spatial filter P-FF where the beam was spatially filtered and its parameters were matched with that of the input window of the spatial light modulator SLM. The resolving power of the spatial modulator was XXX, the pixel size is XXX that is sufficiently for shaping a forked holographic grating for vortex beams with a topological charge of no lower than $m > 70$ and a radial number $n > 50$. The sizes of the holographic grating were controlled in such a way as to ensure the corresponding radius $R$ of the circular hard-edged aperture and the required vortex topological charge $m$ and radial number $n$. Then the beam was split by the beam splitter BS into two arms. The perturbed beam in the first arm was focused by a spherical lens $L2$ ($f_{sh} = 10\,cm$) onto the input window of the charge-coupled device (CCD1) located at the lens focal plane where the intensity distribution processing and mode sorting were performed. The second auxiliary arm made it possible to check the topological charge $m$ of the perturbed beam and its orbital angular momentum (OAM). For this, a cylindrical lens CL ($f_{cyl} = 20\,cm$) was placed into the path of the beam, and the topological charge and OAM measuring were carried out by the standard technique described in Ref.[40,53].

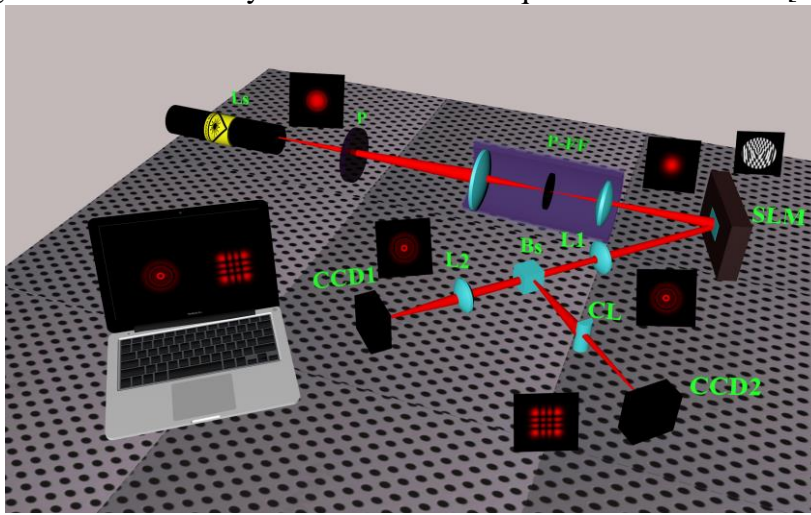

*Fig. 3. Scatch of the experimental setup for real-time measuring the vortex and the OAM spectrum, P – polarizer, FF – space light filter, SLM – space light modulator, L1, L2 – spherical lenses with a focal length $f_{sh}$, BS – beam splitter, CL – cylindrical lent with a focal length $f_{cyl}$, charge – coupled device – CCD1,2*

Let us proceed with the sorting problem in more detail. A sketch of the algorithm for processing the intensity distribution $\mathfrak{I}_{m,n}(r,\varphi)$ and mode sorting is shown in Fig. 5. The algorithm is equipped with a feedback circuit for monitoring the mode sorting and made it possible to sort LG beams with a minimal error of sampling the mode states. The final result of the algorithm is of displaying the squared amplitudes $C_{m,n}^2(k)$ and amplitudes $C_{m,n}(k)$, as well as the information entropy $H(R)$ [46] that serves as a measure of uncertainty inserted by the perturbation.

The algorithm works as follows. First, the algorithm digitizes the intensity distribution $\mathfrak{I}_{m,n}(r,\varphi)$. The next step is a set $\mathcal{M}_{p,q} = M_{p,q} \mathfrak{I}_{m,n}$, as shown in Fig. 5. Next, the image is processed and the integral of the intensity moment $J_{p,q}$ is calculated for each index $p$ and $q$. The results obtained completely determine the vector $\mathbf{J}$ on the right side of the equation.

The next step is to calculate elements of the matrix $\hat{M}$ in Eq.(16) and its determinant $\det \hat{M}$. If this is not zero ($\det \hat{M} \neq 0$), the inverse matrix $\hat{M}^{-1}$ is determined but if the determinant vanishes ($\det \hat{M} = 0$) then the algorithm returns a process to specify the form of the moment function. $M_{p,q}$. Further, the results obtained are substituted into the equation (18) and the squared amplitudes $X_s$ and cross amplitudes $Y_s$ are calculated. The components $X_s$ of the vector $\mathbf{X}$ make it possible to plot the spectrum of squared amplitudes $C_{m,n}^2(k)$. The spectrum $C_{m,n}^2(k)$ allows one to calculate the informational entropy $H(m,n)$ for a given perturbation $R$ value (see below). The components $Y_s$ of the vector $\mathbf{X}$ allow one to determine the signs of cross amplitudes $2C_j C_k$. If the amplitude of the initial beam $C_{m,n}$ is known, then a special program calculates the signs of all other amplitudes $\sigma_{m,n,k} = sign(C_{m,n,k})$ and plots the amplitude spectrum $C_{m,n}(k) = \sigma_{m,n,k}\sqrt{C_{m,n}^2(k)}$. The spectra $C_{m,n}^2(k)$, $C_{m,n}(k)$ and the informational entropy $H(m,n)$ for the given aperture radius $R$ are stored in computer memory.

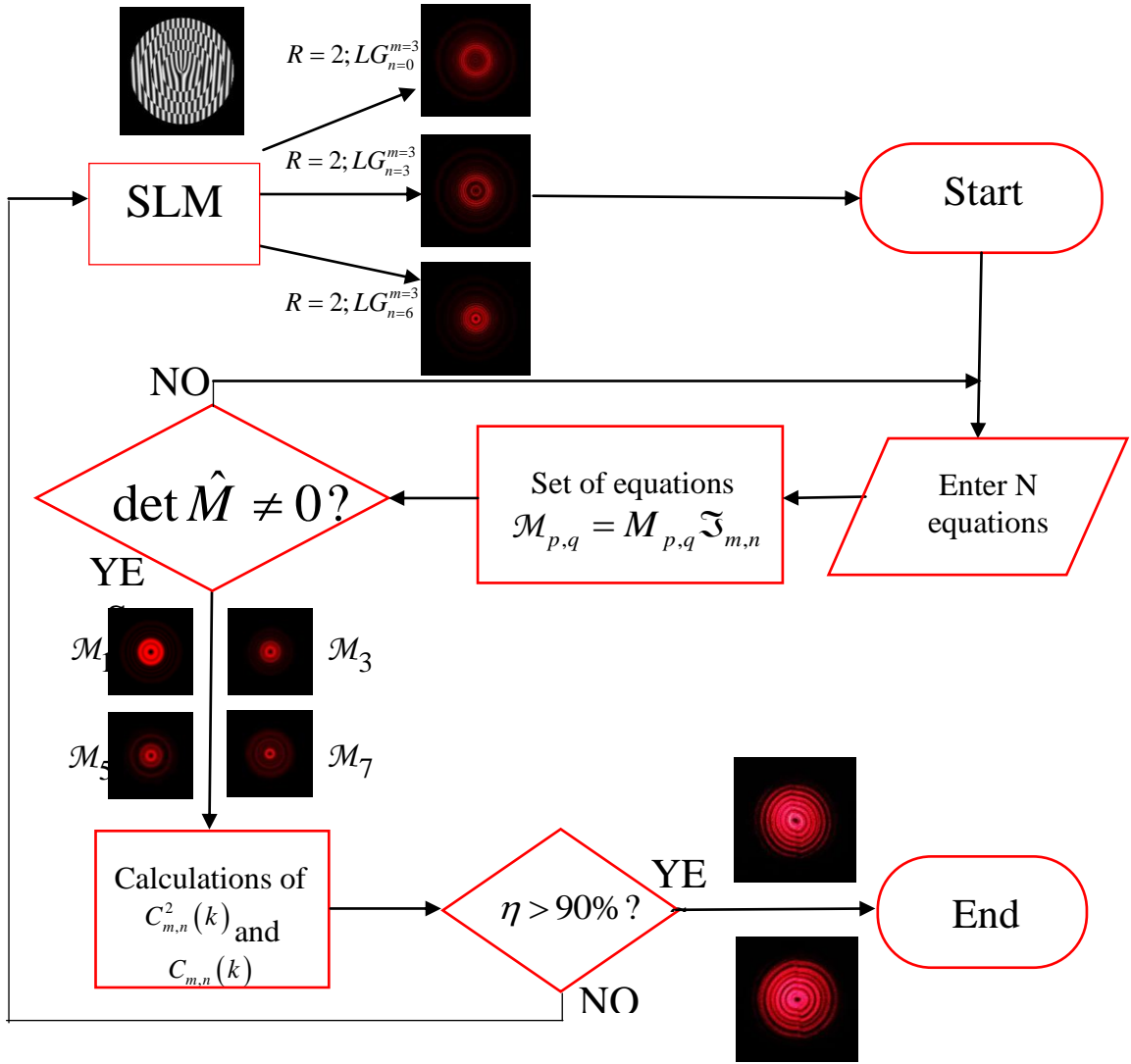

*Fig.5 Sketch of the intensity distribution processing and LG mode sorting by radial numbers*

Based on the amplitude spectrum $C_{m,n}(k)$, the field function $\Psi_{m,n}^{\exp}$ in Eq.(6) is calculated then turn on the feedback circuit and a binary holographic grating was formed on the working element of the SLM modulator in accordance with the expression

$$T_{m,n}^{\exp} = signum\left[\cos\left(\arg \Psi_{m,n}^{\exp} - Q\, r\, \cos\varphi\right)\right], \qquad (19)$$

where $Q$ is a scale parameter. The holographic grating restored a perturbed beam so that its intensity distribution $\mathfrak{I}_{m,n}^{exp}$ was digitized. A comparison of the two intensity distributions $\mathfrak{I}_{m,n}$ and $\mathfrak{I}_{m,n}^{exp}$ was specified by the correlation degree

$$\eta = \int_0^\infty \mathfrak{I}_{m,n}(r)\mathfrak{I}_{m,n}^{exp} r dr / (\int_0^\infty \mathfrak{I}_{m,n}(r) r dr \int_0^\infty \mathfrak{I}_{m,n}^{exp}(r) r dr). \tag{20}$$

If the correlation degree is less than the optimum value $\eta < 0.9$ then the sorting process is further refined. Otherwise ($\eta \geq 0.9$), the spectrum of squared amplitudes $C_{m,n}^2(k)$, the spectrum of amplitudes $C_{m,n}(k)$ and informational entropy are displayed on a computer monitor. This completes the algorithm for the digital sorting of LG modes by the radial number $k$.

The experimental amplitude spectra $C_{m,n}(k)$ shown in Fig. 6 illustrate the characteristic features of the of the perturbed LG beams with the same aperture radius $R=2$ and topological charge $m=3$ but different initial radial numbers $n=0$, $n=3$, $n=6$ and $n=8$. The vortex beam perturbation with $n=0$ in Fig. 6a does not practically make significant changes in the vortex structure apart from appearing two adjacent modes with small amplitudes but a phase shift by $\pi$. The spectral composition pattern changes as the initial radial number grows. A beam with $n=3$ (Fig.6b) already gives rise a broad mode range so that the initial beam energy are also redistributed among modes with radial numbers $n=0, 6, 8$ and a binary phase set $0$ and $\pi$. The same pattern manifests itself for the rest LG beams with $m=6$ and $8$ in Fig.6c and d.

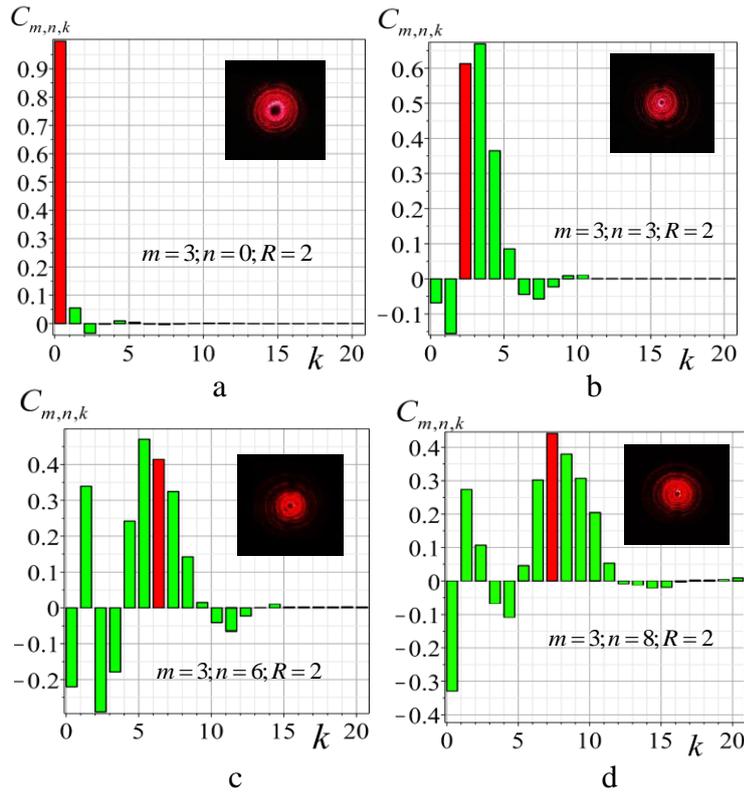

Fig. 6 The aamplitude spectrum $C_{m,n}(k)$ of a LG beam with a topological charge $m=3$ perturbed by a circular aperture $R=2$ for four different initial radial numbers (a) $n=0$, (b) $n=3$, (c) $m=6$ and (d) $m=8$. Callouts: intensity distributions $\mathfrak{I}_{m,n}(r)$ of the corresponding perturbed LG beams. The amplitudes of modes with initial radial numbers $m=0,3,6$ and $8$ are colored red

A more general picture of the energy contribution $C^2_{m,n,k}(R)$ of the initial beams in the states $(m,n)$ to the spectral component $k=0$ at various radii $R$ is shown in Fig. 7.

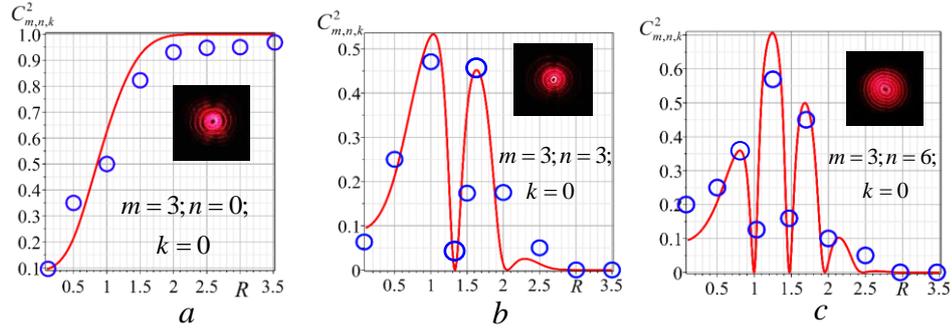

*Fig. 7 Contribution of perturbed LG beams with the initial topological charge $m=3$ and different radial numbers (a) $n=0$, (b) $n=3$ and (c) $n=6$ to the $k=0$ component of the squared amplitudes spectrum $C^2_{m,n,k}(R)$ for different radii $R$ of the circular aperture. Callouts: experimental intensity distributions of the corresponding perturbed LG beams at the aperture radius $R=1$. Solid lines – theory, circulars - experiment*

The monotonic theoretical curve and the accompanying experimental circlets of the energy redistribution $C^2_{m,n,k}(R)$ in Fig. 7a show lowering a contribution of the excited modes from $C^2_{m,n,k} = 0.1$ to $C^2_{m,n,k} = 0.1$ to the initial beam $n=0$ with increasing the aperture radius $R$. The monotonic nature of the energy redistribution gives way to fast oscillations of the theoretical curve as the radial number $n$ increases in Fig. 7b and c.

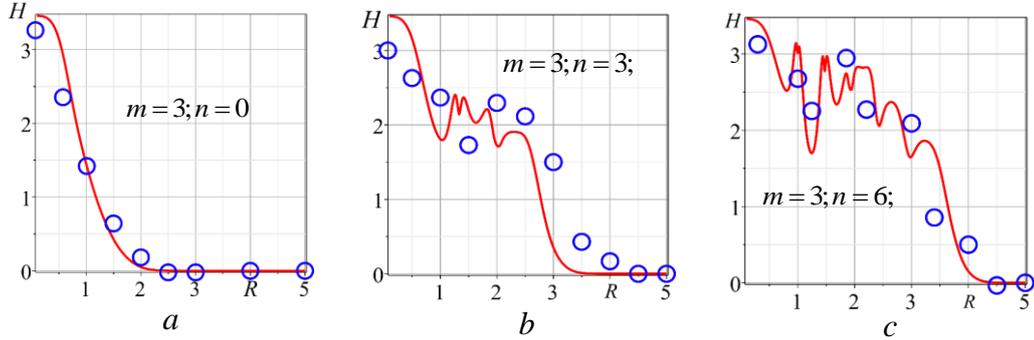

*Fig. 8 Dependency of the informational entropy $H$ on the circular aperture radius $R$ for different initial LG beams with a vortex topological charge $m=3$ and different radial numbers (a) $n=0$, (b) $n=3$, (c) $n=6$. Solid lines – theory, circulars - experiment*

It is interesting to note that the maxima of the contribution of the initial beam to the $k=0$ mode $C^2_{3,3,0}|_{max}$ fall at the zeros of the Laguerre polynomial $L^2_3(2r^2) = 0 \rightarrow r_{max} = R_0 \approx (1.035;\ 1.63;\ 2.3)$ while the zero contribution $C^2_{3,3,0} = 0$ corresponds to the extrema of the polynomial $\frac{d}{dr}L^2_3(2r^2) = 0 \rightarrow r_0 = R_{max} \approx (1.33,\ 2.06)$. A similar situation is also observed for the initial state $(3, 6)$. The energy transfer from higher harmonics $(m,n)$ to the vortex mode with stops when the aperture radius reaches the optimal spot size of the initial beam $R_{cr} = r_{spot} = \sqrt{2n+m+1}$ [54]. As can be seen from positions of the experimental points in Fig. 7, we observe a slight disagreement of the theory and experiment. Such a mismatch, in our opinion, is caused by the fact that the intensity distribution $\Im_{\exper}(r,\varphi)$ has not completely axially symmetric form, i.e. dependence on the azimuthal coordinate $\varphi$ arises, whereas our theoretical predictions do not take it into account. The indicated remark means the heightened requirements to the adjustment of the experimental setup. However, as we show below, the obtained mismatch does not exceed the requirements for the correlation degree $\eta$ of the beam sorting results.

Increasing the state disorder in the perturbed LG beam can be found by measuring the informational entropy $H_{m,n}$ (the Shannon's entropy) [46]. To determine the entropy $H_{m,n}$ in our case we note at first that the normalized squared amplitude $C_n^2 \in (0,1)$ in the expansion (2) can be treated as a conditional probability $P(k/n)$ of finding out a vortex beam in the state $|k\rangle$ among $N$ states, provided that the external perturbation $R$ affected the vortex state $|n\rangle$ i.e. $P(k/n) = C_{m,n}^2(R,k)$. Such an approach to counting disorder in the vortex states can be used in the Shannon formula [55] written in our case as

$$H_{m,n} = -\sum_{k=0}^{N} P(k/n) \log_2 P(k/n) = -\sum_{k=0}^{N} C_{m,n}^2(R,k) \log_2 C_{m,n}^2(R,k) > 0 \quad (21)$$

and measured in bits. On the one hand, entropy $H_{m,n}$ can be treated as a measure of uncertainty caused by the beam perturbation but, on the other hand, it can be interpreted as a measure of new information that appeared in the LG beam after the perturbation. It is important to note that informational entropy $H_{m,n}$ is specified only by the spectrum of squared mode amplitudes $C_{m,n,k}^2$. Figure 8 illustrates increasing uncertainty in the initial states $(3,0)$, $(3,3)$ and $(3,6)$ in the form of $H_{m,n}(R)$ as a result of truncating the LG beam with a circular hard edged aperture. All three beam states are characterized by the same type of the entropy evolution: high values of the entropy $H_{m,n}$ correspond to small aperture radii $R$ and lowering to zero at large radii. However, if the beam entropy in the state $(3,0)$ experiences a monotonic decrease then the entropy lowering of the $(3,3)$ and $(3,6)$ states are accompanied by fast oscillations that have been experimentally recorded.

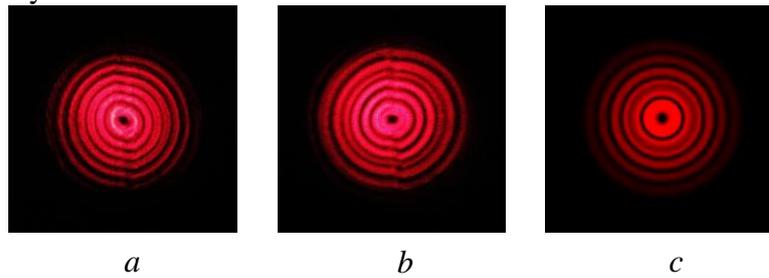

*a*          *b*          *c*

*Fig.9 Intensity distributions $\Im(r,\varphi)$ of the vortex beam with a topological charge $m=3$ and a radial number $n=6$ perturbed by a circular aperture with a radius $R=2.5$: (a) the initial perturbed beam, (b) the beam restored by the sorting mode beams, the correlation degree is $\eta=0.94$, (c) theory*

The final step of the digital sorting process is shaping a beam mixture using amplitude spectra $C_{m,n,k}$ in Fig.6 and 7, engineering a digital holographic grating at the liquid crystal cell of the SLM modulator, and recovering a combined beam. The result of such a sorting is shown in Fig. 9. The first two images in Fig. 9a and b refer to the initial beam in the state $(3,6)$ and to the beam engineered of the sorted modes, respectively while the image in Fig. 9c displays the theoretical plotting. The correlation degree of the first two images is $\eta = 0.94$ that corresponds thoroughly to the conditions of the digital beam sorting.

## 4. Sorting degenerate beam states

In information transmission and data processing systems, one utilizes a mixture of beams with different initial amplitudes and phases. We simplify the problem and choose a mixture of monochromatic LG beams with real amplitudes $A_n$ without initial phases and perturb the combined beam by a circular hard-edged aperture. Similar to that of a single beam perturbation in Sec.3 (see Eq. (11) ) we write the complex beam amplitude in the form

$$\Psi_m(r,\varphi,z) = \frac{e^{-\frac{r^2}{\sigma(z)}+im\varphi}}{\sigma(z)} \sum_{n=0}^{M} A_{m,n} \sum_{k=0}^{N} C_{m,n,k} \left(\frac{r}{w(z)}\right)^{|m|} L_k^{|m|}\left(2\frac{r^2}{w^2(z)}\right) e^{i\Gamma_{m,k}(z)}, \quad (22)$$

where $M$ is a total number of initial LG beams in the vortex mixture. We see that each $k$–th component in the amplitude spectrum $C_{m,k}$ corresponds to $M$ LG mode beams with amplitudes $A_{m,n}$, i.e. each component has $M$–fold degeneracy. In order to remove the degeneracy it is necessary to subject a combined beam to an appropriate transform. As our preliminary estimations showed, one of the possible approaches is to subject the beam to an additional perturbation using a sectorial aperture, as was done in Ref. [43]. In contrast to a circular hard edged aperture, the sectorial aperture does not change the radial number of the perturbed beam $(n = const)$ but generates a broad range of vortex beams with different topological charges $m$. The vortex modes of such a perturbed beam are specified by the sum $|m|+n$ and difference $||m|-n|$ of the topological charge and the radial number. However, this approach requires detailed study and is the subject of a separate consideration. In this paper, we make use of an alternative approach that will be clarified hereinafter.

An example of our computer simulation of a complex perturbed beam that contains M = 15 original LG modes is shown in Fig. 10.

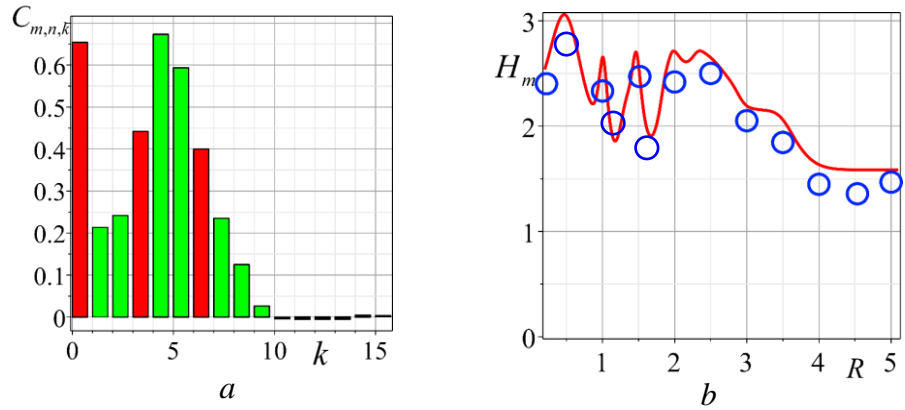

*Fig.10 (a) Amplitude spectrum $C_{m,n}(k)$ of the complex beam of three LG beams with $m = 0$, $n_1 = 0, n_2 = 3, n_3 = 6$, $R = 2$ (red collar); (b) entropy $H_m(R)$, thin line – theory, circlets – experiment.*

First, we measure the amplitude spectrum $C_{m,n,k}$ with respect to a certain initial radial number $n \in \{M\}$ in accordance with the algorithm discussed in the previous section (see Fig.10a). This beam state is described in terms of the amplitude distribution $C_{m,n,k}(R)$ for different aperture radii $R$ shown in Fig. 10b and the entropy $H_m(R)$ in Fig. 10c. Since we know that the beam contains $M=15$ initial LG modes then each k-th spectral component of the perturbed beam contains $M=15$ secondary modes $C_{m,k} = \sum_{n=0}^{M} A_n$. In order to decipher the obtained spectrum in Fig. 10a, one needs to hold $M$ keys. Such keys are the amplitude spectra $C_{n.m.k}(R)$ of LG beams for the given aperture radius $R$, similar to that is done for analyzing radiation spectra of various optical sources (see e.g. Ref.[56]). Of particular interest is the Shannon's entropy of a complex beam. According to Shannon (see e.g. p. 8 in Ref.[46]), the entropy of a complex signal is calculated as

$$H_m = \sum_{n=0}^{M}\sum_{k=0}^{N} \tilde{C}^2_{m,n,k} \log_2 \tilde{C}^2_{m,n,k} \quad (bit) . \tag{24}$$

where $p_n = A^2_{m,n}$ is a probability to meet n-th mode in the initial complex beam and $\tilde{C}^2_{m,n,k} = p_n C^2_{m,n,k}$ stands for renormalized squared amplitudes $C_{n.m.k}$. In other words, each k-th component in the sum (11) makes a contribution to each n-th component of the sum (24) with the probability $p_n$. Therefore, the entropy $H_m$ of a complex beam can be measured similar to the technique considered in the previous section and then compared with computer simulation according to Eq. (24)

For experimental sorting of LG modes of a complex beam, we focus on the combination of three beams with the same unit amplitudes and radial numbers. Thus, the probability of encountering the LG mode in a beam mixture is $p_n = 1/3$. Figure 10a shows the amplitude spectrum $C_{mn}(k)$ of the complex beam perturbed by a circular aperture of the radius $R=2$. The mode amplitudes of the unperturbed beam are highlighted in red. Each mode beam in the a complex beam spectrum makes its contribution to each line of the complex spectrum, say $C_{3,0} = C_{3,0,0} + C_{3,3,0} + C_{3,6,0}$. For example, using the single spectra of Fig. 6a,b,c, we find $C_{3,0,0} \approx 0.9$, $C_{3,3,0} \approx -0.05$, $C_{3,6,0} \approx -0.22$, so that $C_0 \approx 0,67$ while the experimental result of measuring the complex spectrum in Fig. 10a is $C_{3,0} \approx 0,65$, i.e. the error of two alternative measurements is $\Delta C_{3,0} \approx 0.02$.

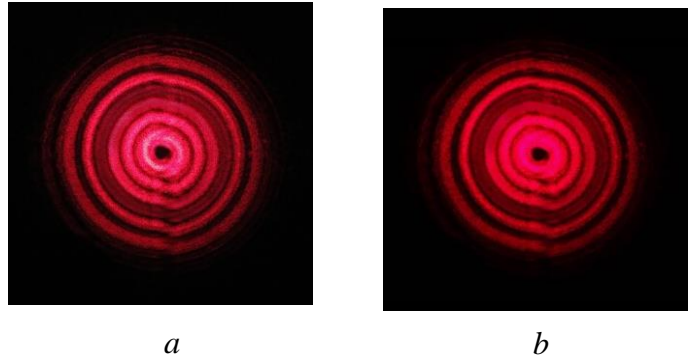

*a*            *b*

Fig.11 (a) Intensity distribution of a complex beam before sorting; (b) intensity distribution of the sorted LG beams with $m=0$, $n_1=0, n_2=3$, $n_3=6$, $R=2$. The correlation degree of the images is $\eta = 0.92$

The measure of uncertainty inserted by the perturbation into the complex beam is specified by the entropy $H_m$, whose dependence on the radius $R$ of the aperture is shown in Fig. 10b. he theoretical curve in the figure first shows fast entropy oscillations that is replaced by an asymptotic tendency to $H_m \approx 1.584962501$ after reaching the critical radius $R_{cr} = 4$. Indeed, as the critical radius of the aperture is reached the value $R_n = \sqrt{m+2n+1}$, the local entropy $H_{m,n}$ of each mode reaches a critical value $H_{m,n} \rightarrow -1/3 \log_2 3$ and does not change further. After the critical radius reaches a spot radius $R_{cr} = 4$ for a mode with a higher radial number $n = 6$, the entropy of the mixture tends to $H_m \rightarrow -3 \cdot (-1/3 \log 3) \approx 1.584962501$. Experimental points follow theoretical predictions with slight deviations. A rather large uncertainty $H_m$ in a complex beam inserts a perceptible error in the mode sorting. Figure 11 shows two images of the complex beams. The intensity distribution before sorting the complex beam in Fig. 11a was compared with the intensity distribution of the complex beam in Fig. 11b obtained by superposition of the sorted out the LG modes. The correlation degree of the intensity distributions was $\eta \approx 0.92$ that barely exceeds the critical value $\eta_{cr} = 0.9$. This indicates the need to improve the experimental technique and the mode sorting method.

## 5. Conclusions

Thus, the results of our studies lie in developing and implementing the basic principles of digital sorting the Laguerre-Gauss modes by radial numbers both for a non-degenerate and a degenerate state of a vortex beam subject to perturbations in the form of a hard-edged aperture of variable radius. The digital sorting of LG beams by the orthogonal basis involves the use of higher-order intensity moments, and subsequent scanning of the modulated beam images at the focal plane of a spherical lens. As a result, we obtain a system of linear equations for the squared mode amplitudes and the cross amplitudes of the perturbed beam. The solution of the equations allows one to determine the amplitudes of each LG mode and restore both the real mode array and the combined beam as a whole.

First, we developed a digital sorting algorithm, and then two types of vortex beams were experimentally studied on its basis: a single LG beam and a composition of single LG beams with the same topological charges $m$ (azimuthal numbers) and different radial numbers $n$. The beam was perturbed by means of a circular hard-edged aperture with different radii R. As a result of the perturbation, a set of secondary LG modes with different radial numbers k is appeared that is characterized by an amplitude spectrum $C_{m,n}(k)$. The spectrum obtained makes it possible to restore both the real array of LG modes and the perturbed beam itself with a degree of correlation not lower than $\eta = 0.94$. As a measure of uncertainty induced by the perturbation we measured the informational entropy (Shannon's entropy).

The perturbation of a complex beam led to appearance of a degenerate spectrum of amplitudes, when a single radial number $k$ in the amplitude spectrum $C_{m,n}(k)$ corresponds to a set of $M$ perturbed LG modes where $M$ is a number of modes in the initial vortex beam. To decrypt and then sort the modes of a degenerate beam, it is necessary to know M keys. As such keys, we chose the amplitude spectra $C_{m,n}(k)$ of non-degenerate perturbed beams for given radii $R$ and initial radial numbers $n$. Measurements showed that the informational entropy of the complex beam increases substantially. This affected the sorting quality of real restored beams in comparison with that of single perturbed beams since the correlation degree decreased to $\eta = 0.92$ but exceeded the optimal correlation degree $\eta_{opt} = 0.90$.

The considered new technique of digital beam sorting makes it possible to significantly simplify the existing devices for sorting beams by radial numbers, known to the authors, since it removes a number of interferometric elements from the optical devices together with auxiliary mechanical and optoelectronic gadgets. Besides, employment of degenerate perturbed beams allows one to use of new quantum key distributions in cryptography, optical communication and data processing systems.

# References

1. L. Allen, M. Padgett. Introduction to Phase-Structured Electromagnetic Waves. Chapter 1 in Structured Light and Its Applications: An Introduction to Phase-Structured Beams and Nanoscale Optical Forces. Academic Press. Elsevier, 2008

2. L. Allen, M. W. Beijersbergen, R. J. C. Spreeuw, J. P. Woerdman. Orbital angular momentum of light and the transformation of Laguerre-Gaussian modes, Phys. Rev. A 45, 8185 (1992)

3. K.Y. Bliokh, F.J. Rodríguez-Fortuño, F. Nori, A.V. Zayats. Spin-orbit interactions of light. Nature Photonics 9 (12), 796-808 (2015)

4. K.Y. Bliokh, D Smirnova, F Nori. Quantum spin Hall effect of light. Science 348, 1448-1451 (2015)

5. T, Fadeyeva, A. Rubass, Yu. Egorov, A Volyar, G Swartzlander Jr. Quadrefringence of optical vortices in a uniaxial crystal. J.Opt.Soc.Amer. A 25 (7), 1634-1641 (2008)

6. T.A. Fadeyeva, A.F. Rubass, A.V. Volyar. Transverse shift of a high-order paraxial vortex-beam induced by a homogeneous anisotropic medium. Physical Review A 79 (5), 053815 (2009)

7. K.T Gahagan, G. A Swartzlander. Optical vortex trapping of particles. Opt. Lett. 21(11), 827-829 (1996)

8. V. Shvedov, A. R. Davoyan, C. Hnatovsky, N. Engheta, W. Krolikowski, "A long-range polarization-controlled optical tractor beam," Nature Photon. 21, 26335–26340 (2014)

9. Optical Tweezers, edited by M. J. Padgett, J. Molloy, and D. McGloin (Chapman and Hall, London, 2010).

10. F. Tamburini, Bo Thidé, G. Molina-Terriza, G. Anzolin. Twisting of light around rotating black holes. Nat. Phys. 7, 195 (2011)

11. G. Foo, D. M Palacios, G. A Swartzlander. Optical vortex coronagraph. Opt. Lett. 30(24), 3308-3310 (2005)

12. J. Wang, J.-Y. Yang, I. M. Fazal, N. Ahmed, Y. Yan, H. Huang, Y. Ren, Y. Yue, S. Dolinar, M. Tur, and A. E. Willner, Terabit free-space data transmission employing orbital angular momentum multiplexing. Nat. Photon. 6, 488-498 (2012)

13. N. Bozinovic, Yang Yue, Yongxiong Ren, M. Tur, P. Kristensen, Hao Huang, A.E Willner, S. Ramachandran. Terabit-scale orbital angular momentum mode division multiplexing in fibers. Science 340 (6140), 1545-1548 (2013).

14. N. Uribe-Patarroyo, A. Fraine, D. S. Simon, O. Minaeva, A. V. Sergienko, Object identification using correlated orbital angular momentum states, Phys. Rev. Lett. **110**, 043601 (2013)

15. C.J.R. Sheppard. Fractal model of light scattering in biological tissue and cells. Opt. Lett., 32(2), 142-144 (2007)

16. M.V. Berry, M.R. Dennis. Knotted and linked phase singularities in monochromatic waves. Proc. R. Soc. Lond. A **457**, 2251–2263 (2001)

17. J. Leach, M. Dennis, J. Courtial, M. Padgett, M. Knotted threads of darkness. Nature 432, 165 (2004).

18. E. Karimi and E. Santamato. Radial coherent and intelligent states of paraxial wave equation Opt. Lett. *37*, 2484-2386 (2012)

19. W. N. Plick, M. Krenn. Physical meaning of the radial index of Laguerre-Gauss beams. Phys Rev. A 92(6) 063841 (2015)

20. E. Karimi, D. Giovannini, E. Bolduc, N. Bent, F. M. Miatto, M. J. Padgett, R. W. Boyd. Exploring the quantum nature of the radial degree of freedom of a photon via Hong-Ou-Mandel interference. Phys. Rev. A, 89, 013829 (2014)

21. E. Karimi, R. W. Boyd, P. de la Hoz, H. de Guise, J. Rehˇa´cek, Z. Hradil, A. Aiello, G. Leuchs, L. L. Sanchez-Soto. Radial quantum number of Laguerre-Gauss modes. Phys. Rev. A, 89, 063813 (2014)

22. A. Mair, A. Vaziri, G. Weihs, A. Zeilinger. Entanglement of Orbital Angular Momentum States of Photons. Nature, 412, 313 (2001)

23. Yiyu Zhou, M. Mirhosseini, S. Oliver, Jiapeng Zhao, S. M. H. Rafsanjani, M. P. J. Lavery, A. E. Willner, R. W. Boyd. Using all transverse degrees of freedom in quantum communications based on a generic mode sorter. Opt. Express, 27(7),10383-10394  (2019)

24. M. Malik, M. Erhard, M. Huber, M. Krenn, R. Fickler, A. Zeilinger. Multi-photon entanglement in high dimensions. Nature Photonics, 10(4), 248-256, (2016)

25 V. A. Soifer, M. A. Golub. Laser beam mode selection by computer-generated holograms» (Boca Raton, CRC Press, 1994)

26. S.N. Khonina, V.V. Kotlyar, V.A. Soifer, P. Pääkkönen, J. Simonen, J. Turunen. An analysis of the angular momentum of a light field in terms of angular harmonics, Journal of Modern Optics, 48:10, 1543-1557 (2001)

27. S. N. Khonina, V. V. Kotlyar, R. V. Skidanov, V. A. Soifer, P. Laakkonen, J. Turunen, and Y. Wang, "Experimental selection of spatial Gauss-Laguerre Modes," Opt. Mem. Neural. Networks 9(1), 73–82 (2000)

28. S. N. Khonina, V. V. Kotlyar, V. A. Soifer, P. Paakkonen, J. Turunen. Measuring the light field orbital angular momentum using DOE . Optical Memory and Neural Networks · January 2001

29. M. Hiekkamäki, S. Prabhakar, R. Fickler. Near-perfect measuring of full-field transverse-spatial modes of light. Opt. Express, 27(22), 31456- 31464 (2019)

30. G. Labroille, B. Denolle, Pu Jian,, P Genevaux, N. Treps, J.-F. Morizur. Efficient and mode selective spatial mode multiplexer based on multi-plane light conversion. Opt. Express, 22(13), 15599 15607 (2014)

31. M. P. J. Lavery, G. C. G. Berkhout, J. Courtial, M. J. Padgett, Measurement of the light orbital angular momentum spectrum using an optical geometric transformation, J. Opt. **13**, 064006 (2011)

32. A. D'errico**,** R**.** D'amelio**,** B. Piccirillo**,** F. Cardano, L. Marrucc**.** Measuring the complex orbital angular momentum spectrum and spatial mode decomposition of structured light beams. Optica, 4, 1350-1357 (2017)

33. J.M. Andersen, S.N. Alperin, A.A. Voitev, W.G. Holtzmann, J.T. Gopinath, M.E. Simens. Characterizing vortex beams from a spatial light modulator with collinear phase-shifting holography. Applied Optics, 58, 404-409 (2019)

34. E. Karimi, E. Santamato. Radial coherent and intelligent states of paraxial wave equation. Opt. Lett. 37(13), 2486-2486 (2012)

35. Robert Fickle, M. Ginoya, R. W. Boyd. Custom-tailored spatial mode sorting by controlled random scattering. Phys. Rev. B, 161108 (2017)

36. Xuemei Gu, M. Krenn,  M. Erhard, A. Zeilinger. Gouy Phase Radial Mode Sorter for Light: Concepts and Experiments. Phys. Rev. Lett., 120, 103601 (2018)

37. Y. Zhou, M. Mirhosseini, Dongzhi, Jiapeng Zhao, S. M. H. Rafsanjani, A. E. Willner, R. W. Boyd. Sorting Photons by Radial Quantum Number.  Phys, Rev. Lett., 119**,** 263602 (2017)

38. D. Fu, Y. Zhou, R. Qi, S. Oliver, Y. Wang, S.M.H. Rafsanjani, J. Zhao, M. Z. Shi, P. Zhang, R.W.Boyd. Realization of a scalable Laguerre–Gaussian mode sorter based on a robust radial mode sorter. Opt. Express, 26(25), 33057- 33065 (2018)

39. D. Mendlovic, H. M. Ozaktas, Fractional Fourier transform and their optical implementation I, J. Opt. Soc. Am. A, 10, 1875-1881, 1993

40. A. Volyar, M. Bretsko, Ya. Akimova, Yu. Egorov. Measurement of the vortex spectrum in a vortex-beam array without cuts and gluing of the wavefront. Opt. Lett. **43**(22), 5635-5638 (2018)

 41. A. Volyar, M. Bretsko, Y. Akimova, Y. Egorov. Vortex avalanche in the perturbed singular beams. J. Opt. Soc. Am. A, 36(6), 1064-1071 (2019)

42. A. Volyar, M. Bretsko, Ya. Akimova, Yu. Egorov. Measurement of the vortex and orbital angular momentum spectra with a single cylindrical lens. Applied Optics, 58(21), 5748-5755 (2019)

43. A. Volyar, M. Bretsko, Ya. Akimova, Yu. Egorov. Orbital angular momentum and informational entropy in perturbed vortex beams. Opt. lett.  44(22), 2687-2690  (2019).

44. Jan Flusser, Tomáš Suk and Barbara Zitová. Moments and Moment Invariants in Pattern Recognition. 2009 John Wiley & Sons, Ltd
45. A. Bekshaev, M. Soskin, M. Vasnetsov, "Optical vortex symmetry breakdown and decomposition of the orbital angular momentum of light beams," J. Opt. Soc. Am. A **20**, 1635 (2003)
46. Fransis T.S.Yu., Entropy and information optics. Marcel Dekker, Inc., New York · Basel, 2000.
47. R.G. Glauber. Quantum theory and optical coherence. WILEY-VCH Verlag GmbH & Co. KGaA,Weinheim, 2007
48. L. Mandel, E. Wolf. Quantum coherence and quantum optics. Cambridge University Press, 1995
49. K. L. Pregnell, D. T. Pegg. Measuring the elements of the optical density matrix. Phys. Rev. A, 66, 013810 (2002)
50. M. J. Bastiaans, "The Wigner Distribution Function Applied to Optical Signals and Systems," Opt. Commun.25,26–30 (1978)
51 E. Cagniot, M. Fromager, K. Ait-Ameur. Modeling the propagation of apertured high-order Laguerre–Gaussian beams by a user-friendly version of the mode expansion method. J. Opt. Soc. Amer. A, 27(3), 484-491 (2010)
52 A. P. Prudnikov, Y. A. Brychkov, and O. I. Marichev, Integrals and Series, Special Functions (Gordon and Breach, 1986)
53 V. V. Kotlyar, A. A. Kovalev, A. P. Porfirev. Astigmatic transforms of an optical vortex for measurement of its topological charge. App, Oprics, 2017, 56, 4095-4104
54. R. L. Phillips, L. C. Andrews, Spot size and divergence for Laguerre Gaussian beams of any order, Appl. Opt. **22**, 643–644 (1983)
55. C. E. Shannon, A Mathematical Theory of Communication, Bell Syst. Tech. J.**,** vol. 27, 379–423, 623–656 (1948).
56. J. Michael Hollas. Modern spectroscopy, John Wiley & Sons Ltd, 2004